\pdfoutput=1
\documentclass[letterpaper]{article}
\usepackage{aaai2026}
\usepackage{times}
\usepackage{helvet}
\usepackage{courier}
\usepackage[hyphens]{url}
\usepackage{graphicx}
\usepackage{natbib}
\usepackage{caption}
\usepackage{booktabs}

\usepackage{adjustbox}
\usepackage{listings}
\lstdefinestyle{tmpl}{basicstyle=\ttfamily\scriptsize,breaklines=true,
  columns=fullflexible,frame=single,framesep=4pt,xleftmargin=2pt,
  aboveskip=3pt,belowskip=0pt,keepspaces=true}

\nocopyright
\title{Fine-tuning LLMs for Tourist Trajectory Prediction using Field Experiment Data\thanks{Accepted at the 2nd Workshop on AI for Urban Planning (AI4UP) at AAAI-26, Singapore, January 2026.}}

\author{
    Tatsuya Amano\textsuperscript{\rm 1,\rm 2},
    Hirozumi Yamaguchi\textsuperscript{\rm 1,\rm 2}
}
\affiliations{
    \textsuperscript{\rm 1}The University of Osaka, Suita, Japan\\
    \textsuperscript{\rm 2}RIKEN Center for Computational Science, Kobe, Japan\\
    t-amano@ist.osaka-u.ac.jp, h-yamagu@ist.osaka-u.ac.jp
}

\begin{document}

\maketitle

\begin{abstract}
Evaluating mobility interventions at tourist destinations requires predicting visitor behavior under varying conditions. Traditional methods struggle because tourist decisions depend heavily on context like weather and fatigue, yet models cannot generalize to unobserved scenarios.
Large Language Models offer a solution by encoding commonsense knowledge about human behavior from pretraining, enabling reasoning about context-dependent decisions, while natural language representation flexibly integrates heterogeneous information. Fine-tuning on local trajectories adapts this general understanding to destination-specific patterns.
We validate this approach using 566 trajectories from Wakayama Castle Park, Japan. Our fine-tuned Llama-3.1-8B achieves 49.1\% next POI accuracy and maintains strong performance on undersampled scenarios like rainy days, demonstrating effective generalization. This establishes LLMs as high-fidelity behavior models for context-dependent tourist prediction, providing groundwork for counterfactual analysis of mobility interventions.
\end{abstract}

\section{Introduction}

Regional tourist destinations worldwide face the challenge of uneven visitor distribution. Popular landmarks attract overwhelming crowds while culturally significant sites nearby remain underutilized, limiting both visitor satisfaction and local economic development \cite{UNWTO2018Overtourism}. This imbalance often stems from physical barriers that appear minor on maps but significantly impact actual visitor behavior. Walking distances become substantial obstacles for elderly visitors and families with young children. The spatial configuration of attractions, combined with limited awareness of alternative sites, creates concentrated flows that stress popular locations while leaving valuable resources underexploited.

To address these circulation challenges, many destinations have begun deploying innovative mobility solutions. Electric shuttles, bike-sharing systems, and autonomous vehicles promise to connect distributed attractions and redistribute visitor flows \cite{YANG2021104328}. In Japan, the Ministry of Land, Infrastructure, Transport and Tourism has supported 37 Green Slow Mobility pilot programs between 2019 and 2021. Our research team conducted field experiments at Wakayama Castle Park in Wakayama Prefecture, implementing three types of mobility services including Green Slow Mobility vehicles, e-bikes, and walking support devices to enhance circulation from the established southern castle area to newly developed northern facilities.

However, a critical challenge remains. While these mobility interventions require substantial investment, predicting how they will actually change tourist behavior is remarkably difficult. Destination planners need to forecast whether introducing a shuttle service will encourage visitors to explore distant attractions, or whether e-bikes will alter route choices and dwell times. Without reliable predictions, effective planning and investment decisions become nearly impossible. Will a new mobility service successfully redistribute flows, or will visitors simply use it to reach the same popular sites faster? The inability to answer such questions before implementation creates significant uncertainty for destination management.

Traditional approaches to tourist behavior prediction struggle with this complexity. Markov chains assume memoryless transitions that ignore how accumulated experiences shape choices. Deep learning methods achieve higher accuracy but require massive datasets rarely available for individual destinations and struggle to incorporate contextual information like weather or special events that significantly influence mobility patterns. Most critically, these models cannot leverage commonsense knowledge about human behavior. They require explicit observation of each scenario to learn patterns that humans understand intuitively, making them unable to predict how visitors might respond to new mobility options or changed conditions.

This research proposes fine-tuning Large Language Models to predict tourist trajectories by combining destination-specific learning with pre-trained behavioral knowledge. LLMs trained on massive text corpora have encoded rich understanding of human decision-making under various conditions. They understand that rain encourages indoor activities, that families need frequent rest breaks, and how transportation options influence destination choices. By fine-tuning these models on local trajectory data, we combine their general reasoning capabilities with specific knowledge about a destination's layout and patterns. The approach processes contextual information naturally through text, describing weather, fatigue, mobility availability, and preferences without manual feature engineering. This enables prediction even for undersampled scenarios by leveraging commonsense reasoning to interpolate between observations.

We evaluate this approach using field experiment data from Wakayama Castle Park, where we collected detailed trajectories from 566 tourists with rich contextual information. Our fine-tuned Llama-3.1-8B model achieves 49.1\% next-POI prediction accuracy, substantially outperforming traditional baselines while maintaining strong generalization to rare contexts. The model demonstrates robust performance on undersampled scenarios like rainy days and generates coherent multi-step trajectories. These results establish the feasibility of using LLMs as a high-fidelity behavior model. While this paper focuses on predictive accuracy, this model serves as a crucial component for future counterfactual simulations to evaluate mobility interventions, opening new possibilities for destination management.

\section{Related Work}

\subsection{Tourist Behavior and Next POI Prediction}

Research on tourist mobility has focused on Next Point of Interest (POI) prediction, primarily using check-in data from Location-Based Social Networks (LBSNs). This field evolved from classical statistical models like Markov chains to more complex deep learning architectures.

Markov models are constrained by their memoryless assumption and suffer from extreme data sparsity when extended to higher orders, failing to capture long-term preferences \cite{feng2018deepmove}. Economic frameworks such as discrete choice models \cite{Train2009DCM} can explicitly model rational decisions but struggle with the vast state space of POI choices.

Deep learning models such as DeepMove \cite{feng2018deepmove} and KGDAE \cite{10.1016/j.ipm.2023.103369} achieved higher accuracy but face critical limitations.
These include massive data requirements unsuitable for individual destinations, an inability to process contextual text, and poor generalization to unobserved scenarios.

\subsection{LLMs for Behavior Prediction and Simulation}

Recent advances in Large Language Models have demonstrated their potential for behavioral modeling beyond traditional NLP tasks \cite{10.5555/3495724.3495883}.
The Generative Agents framework \cite{Park2023GenerativeAgents} showed that LLM-powered agents could produce believable daily routines and social interactions without explicit programming, maintaining memories, reflecting on experiences, and planning activities using commonsense reasoning rather than predetermined rules.
LLMs have been applied directly to mobility prediction. LLM-Mob \cite{wang2023would} achieved the first zero-shot next location prediction, harnessing the pretrained geographical and social knowledge of GPT-3.5/4. AgentMove \cite{feng2025agentmove} advanced this with an agentic framework for worldwide prediction, outperforming baselines across 12 cities without city-specific retraining.
LLM-MPE \cite{LIANG2024102153} demonstrated prediction under public events using textual descriptions.

However, most research on LLM agents operates within synthetic environments or focuses on zero-shot prediction without fine-tuning on local trajectories.
Validation against fine-grained, real-world behavioral data from live field experiments in specific destinations is limited. It remains unclear whether LLMs can capture the complex, context-dependent decisions of tourists in real-world settings and leverage their commonsense reasoning to generalize to unobserved scenarios within that specific location.

\section{Methodology}

\subsection{Problem Formulation}

We formulate tourist trajectory prediction as conditional sequence generation. Each trajectory $h^{(u)}$ consists of time-ordered POI visits:
$$h^{(u)} = \{(p_1, t_1, a_1, c_1), \ldots, (p_T, t_T, a_T, c_T)\}$$
where $p_i$ denotes the POI, $t_i$ the visit time, $a_i$ the area, and $c_i$ the category. The model learns the conditional distribution:
$$P(h|u,e;\theta) = \prod_{i=1}^{T} P(p_i, t_i, a_i, c_i | u, e, h_{<i}; \theta)$$
where $u$ represents tourist persona (age, gender, group type), $e$ captures environmental conditions (weather, time), and $h_{<i}$ denotes visit history before step $i$.

\subsection{Data Collection and Preprocessing}

\begin{figure}[t]
\centering
\includegraphics[width=0.55\columnwidth]{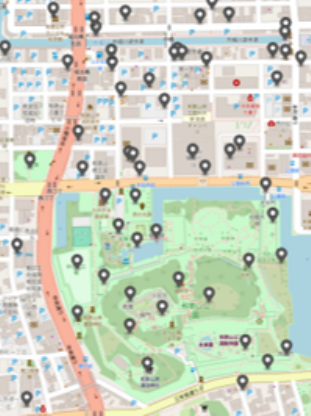}
\caption{Distribution of POIs at Wakayama Castle Park}
\label{fig:wakayama}
\end{figure}

We collected 566 trajectories at Wakayama Castle Park in December 2023 during a Green Slow Mobility pilot program (Figure \ref{fig:wakayama}). The park contains 68 POIs including the castle tower, Momijidani Garden, zoo, museums, restaurants, and rest facilities. Data came from two sources: 87 GPS-tracked participants (1-second intervals) and 479 QR code stamp rally participants who scanned codes at 37 major attractions. We added 31 POIs from OpenStreetMap (cafes, stores, restrooms, hotels) for comprehensive coverage.

Demographics were collected via exit surveys. Missing attributes (18\% of samples) were inferred using GPT-4o based on visit patterns and timestamps. GPS traces underwent stop detection (within 10m for over 60s) and POI matching (nearest within 25m). QR sequences were processed directly after deduplication. Both sources yielded unified representations with POI names, timestamps, areas, and categories, augmented with weather data. The combined dataset averages 7.7 POI visits over 58 minutes, ranging from 2 to 40 POIs and 33 minutes to 5.6 hours.

\subsection{Model Architecture and Text Representation}

We fine-tuned Llama-3.1-8B, extending its vocabulary with special tokens for all POI names, areas, categories, and structural tags. Token embeddings were initialized by averaging constituent subwords from the original tokenizer. Trajectories are encoded as structured text (Figure~\ref{fig:encoding}).
Each visit line contains timestamp, action, area, category, and POI, preserving sequential dependencies with rich contextual information.

\begin{figure}[t]
\centering
\begin{lstlisting}[style=tmpl]
# persona and environment, fixed for the trajectory
<PERSONA>30s, Male</PERSONA><ENV>Sunny day</ENV>
# one line per POI visit, in time order
<time>09:55</time><action>visit</action><area>Castle Tower Area</area><category>Historic Site</category><POI>Wakayama Castle Tenshukaku</POI>
<time>10:25</time><action>visit</action><area>Garden Area</area><category>Garden</category><POI>Momijidani Garden</POI>
...
\end{lstlisting}
\caption{Text representation of a trajectory. Angle-bracketed tags are added to
the vocabulary as special tokens; \texttt{\#} lines are annotations.}
\label{fig:encoding}
\end{figure}

\subsection{Training Procedure}

During training, the model generates complete trajectories given persona and environment via supervised fine-tuning with standard cross-entropy loss. At inference, we provide visit history and prompt for the next POI. Specifically, we input up to the last area tag and generate the subsequent category and POI tags.
We employed QLoRA (Quantized Low-Rank Adaptation) with rank $r=32$ and scaling $\alpha=64$. Crucially, we applied adapters not only to attention projection layers (Query, Key, Value, Output) but also to embedding (embed\_tokens) and output (lm\_head) layers. This inclusion proved essential for learning effective representations of newly added POI tokens.
Training used AdamW optimizer with learning rate $2 \times 10^{-5}$, batch size 8, and ran for 10 epochs. We split 80\% (453 tourists) for training and 20\% (113 tourists) for testing, maintaining tourist-level separation to prevent data leakage.

\section{Experiments}

\subsection{Experimental Setup}

We evaluated 566 tourist trajectories from Wakayama Castle Park with 80\% training (453 tourists) and 20\% test (113 tourists) splits maintaining tourist-level separation. Experiments were conducted on a server with two NVIDIA A6000 GPUs.

Baselines included Markov Models (1st and 5th order), a 2-state Hidden Markov Model, and GPT-4o in zero-shot and fine-tuned configurations using OpenAI's fine-tuning API. We also compared against Llama-3-Swallow-8B \cite{Okazaki:COLM2024}, a Japanese-specialized variant of Llama-3. Since our field site is a Japanese destination, Swallow tests whether Japanese-specific pre-training provides advantages over general multilingual models. For fair comparison, Swallow used identical QLoRA settings to our Llama-3.1 model.

Evaluation metrics included Accuracy@1 for POI and category prediction, and sequence-level metrics (n-gram overlap, BLEU, normalized Levenshtein distance) for coherence assessment.

\subsection{Prediction Performance}

\begin{figure}[t]
\centering
\includegraphics[width=0.9\columnwidth]{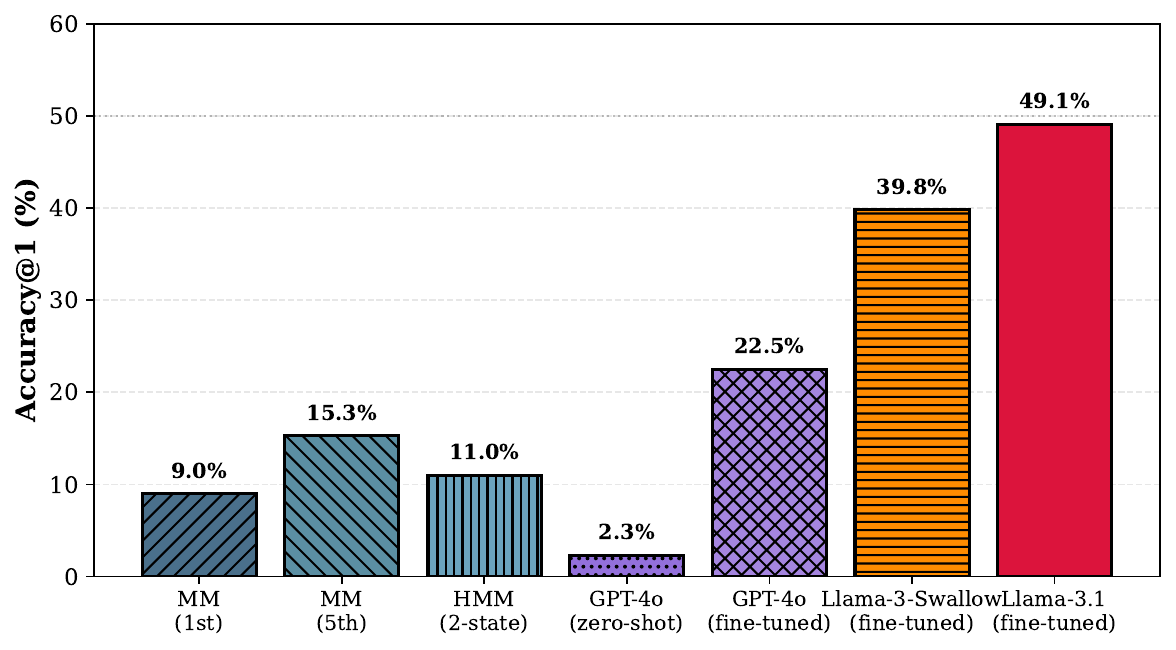}
\caption{Next POI prediction accuracy comparison across methods. }
\label{fig:main_results}
\end{figure}

Figure \ref{fig:main_results} presents prediction accuracy across methods. Statistical methods achieved limited performance: first-order Markov reached 9.0\% (memoryless assumption), fifth-order improved to 15.3\% (but suffered data sparsity with 68 POIs), and Hidden Markov Model achieved 11.0\%.

Zero-shot GPT-4o achieved 18.7\%, showing some generalization from pre-training alone. Fine-tuning via OpenAI's API improved this to 35.2\%, confirming the value of local adaptation though still substantially below our results.

Our fine-tuned Llama-3.1 achieved 49.1\% POI accuracy and 55.4\% category accuracy, substantially outperforming all baselines. Llama-3-Swallow achieved 39.8\% POI and 45.8\% category accuracy. Despite Japanese-specific pre-training, Swallow underperformed Llama-3.1, suggesting that the scale and diversity of multilingual pre-training outweigh language-specific advantages for this task. This implies that general behavioral understanding from massive diverse corpora is more valuable than linguistic specialization when reasoning about physical movement patterns.

The six-point gap between POI and category accuracy suggests that while general visitor intentions are predictable, specific choices among similar options retain inherent uncertainty. The strong performance stems from combining pre-trained commonsense knowledge (e.g., seeking lunch at midday, preferring indoor venues during rain) with destination-specific learning through QLoRA fine-tuning.

\begin{figure}[t]
\centering
\includegraphics[width=0.9\columnwidth]{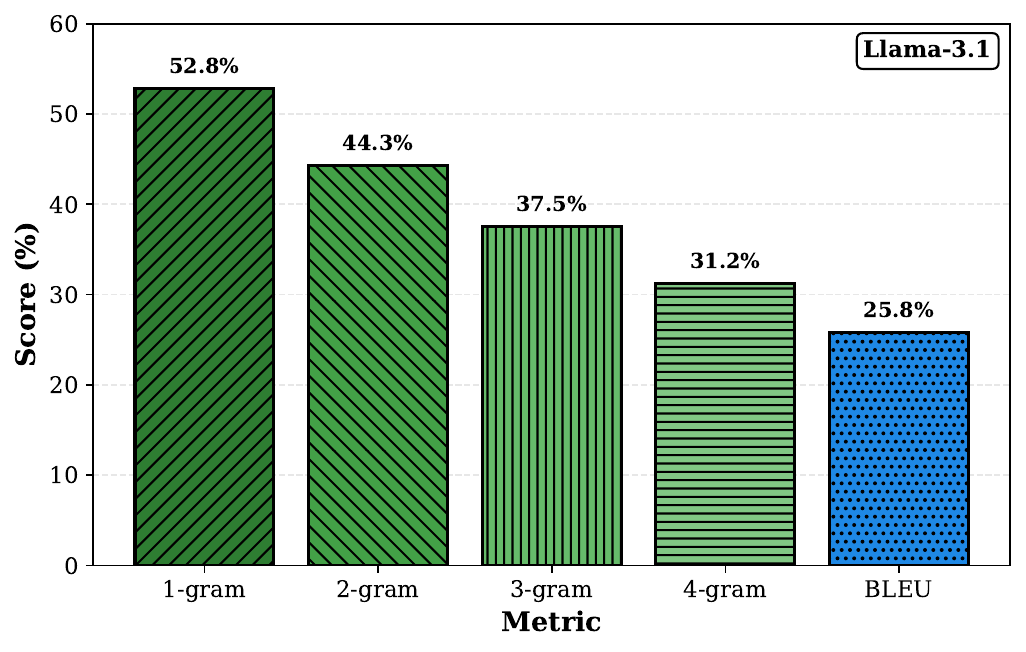}
\caption{POI-Sequence-level evaluation metrics for Fine-tuned Llama-3.1. }
\label{fig:sequence_metrics}
\end{figure}

\subsection{Sequence Generation Quality}

Figure \ref{fig:sequence_metrics} shows sequence-level metrics. Llama-3.1 achieved 31.2\% 4-gram overlap, 25.8\% BLEU, and 47.0\% normalized Levenshtein distance, capturing longer sequential patterns. Generated trajectories exhibit realistic patterns: morning visits to major attractions, midday dining, afternoon rest areas. Average length (7.2 POIs) matches real data (7.7) without explicit constraints. Anomalous outputs occurred in only 5.4\% of generations.

The model generalizes to undersampled contexts. On 12 rainy test samples, our model maintained 41.7\% accuracy versus 8.3\% for Markov models, correctly increasing indoor predictions while reducing garden visits. Lunch-time predictions achieved 62.3\% category accuracy for dining, demonstrating effective temporal adaptation through commonsense reasoning.

\section{Conclusion}

This work demonstrates the feasibility of using fine-tuned Large Language Models for tourist trajectory prediction at regional destinations. Our Llama-3.1-8B model achieved 49.1\% next-POI accuracy on 566 real-world trajectories from Wakayama Castle Park, substantially outperforming traditional statistical and neural baselines. The model successfully leverages pre-trained commonsense reasoning to generalize to undersampled contexts such as rainy days and generates coherent multi-step trajectories that reflect realistic visitor behavior patterns.

This work establishes predictive capability as a necessary foundation for counterfactual generation but does not validate causal claims. Predicting what tourists would do under current conditions differs from estimating behavioral changes under hypothetical interventions. Future work should generate counterfactuals by modifying input conditions (e.g., adding shuttle service availability to the context) and validate predictions against A/B test data or randomized field trials to enable causal evaluation of mobility interventions.

Several limitations remain. Missing persona attributes (18\%) were inferred via GPT-4o, though evaluation on manually verified subsets showed minimal impact. The model handles weather and time variations but cannot predict behavior at entirely new POIs without descriptions or reason about major infrastructure changes. Addressing these may require architectural extensions such as retrieval-augmented generation or few-shot adaptation techniques. Despite these limitations, this work provides a foundation for LLM-based destination management tools that could transform how tourism planners evaluate and optimize mobility interventions.

\section{Acknowledgments}

This work was supported by JST PRESTO Grant JPMJPR2361.

\bibliography{ref}

\end{document}